\pdfoutput=1
\documentclass{ptapap}
\usepackage{amsmath}
\usepackage{amssymb}

\author{Maciej Jab\l o\'{n}ski}[CAMK,UJ]
\author{Aleksandra Kotek}[CAMK,UW]
\author{Miljenko \v{C}emelji\'{c}}[CAMK,AS]
\author{W\l odek Klu\'{z}niak}[CAMK]
\affil[CAMK]{Nicolaus Copernicus Astronomical Center, Polish Academy of Sciences, Bartycka 18, 00--716 Warsaw, Poland}
\affil[UJ]{Astronomical Observatory, Jagiellonian University, ul. Orla 171, 30-244 Kraków, Poland}
\affil[AS]{Academia Sinica, Institute of Astronomy and Astrophysics, P.O. Box 23-141, Taipei 106, Taiwan}
\affil[UW]{Faculty of Physics, University of Warsaw, Pasteura 5, 02--093, Warsaw, Poland}

\title{Towards pseudo-Newtonian black hole jets: comparison of forces}

\begin{document}

\maketitle

\begin{abstract}
We perform resistive and viscous MHD simulations of axial jet launching from
a magnetized thin accretion disk around a supermassive black hole. We
compare the forces in our pseudo-Newtonian simulations for the
Paczyński-Wiita and Kluźniak-Lee potentials. The results will help to find
magnetic field configurations and physical parameters conducive to jet
launching.
\end{abstract}

\section{Introduction}

Accretion onto a supermassive Black Hole in Active Galactic Nuclei is
extensively studied both in observations and in numerical simulations. 
Launching mechanism of outflows and axial jets is still not fully explained. 
We extend our Newtonian non-ideal MHD simulations with a magnetized
accretion disk around a central object to the simulations with a
pseudo-Newtonian gravitational potential, to study the jet launching from
AGNs.

\section{Numerical setup}

We use the PLUTO code \citep{m07} to perform 2D-axisymmetric star-disk
simulations in the complete $[0,\pi]$ half-plane, in resolution
$R\times\theta$=$[125\times100]$ grid cells. The physical domain extending
from [1-50] R$_{\mathrm i}$, where R$_{\mathrm i}$ is set to three
Schwarzschild radii $r_{\mathrm S}=2GM/c^2$.  The equations we solve with
the PLUTO code are the resistive and viscous magneto-hydrodynamic equations:

\begin{align}
\frac{\partial \rho}{\partial t}+\nabla \cdot (\rho \mathbf{v})=0& \nonumber \\
\nabla \cdot \mathbf{B}=0& \nonumber \\
\frac{\partial \rho \mathbf{v}}{\partial t}+\nabla \cdot \left [\rho \mathbf{v} \mathbf{v} +\left(P+\frac{B^2}{8 \pi} \right) \tilde{I}-\frac{\mathbf{B} \mathbf{B}}{4 \pi}-\tilde{\tau} \right] =-\rho \nabla \Psi_{g} \nonumber \\
\frac{\partial E}{\partial t}+\nabla \cdot \left [\left(E+P+\frac{B^2}{8 \pi} \right) \mathbf{v} - \frac{(\mathbf{v} \cdot \mathbf{B})\mathbf{B}}{4 \pi} \right] =-\rho \nabla \Psi_{g} \cdot \mathbf{v} \nonumber \\
\frac{\partial \mathbf{B}}{\partial t}+\nabla \times (\mathbf{B} \times \mathbf{v}+ \eta_m \mathbf{J})=0
\label{eq1}
\end{align}

where $\rho$, P, v, B, $\eta_{\mathrm m}$ are the density, pressure, velocity, magnetic field and
the Ohmic resistivity, respectively. The terms $\tilde{I}$ and $\tilde{\tau}$ are representing the
unit tensor and the viscous stress tensor, respectively. In the energy equation, we assumed that
all the dissipation (viscous and resistive) heating is radiated away, and removed the corresponding
terms in the code. The Paczy\'{n}ski-Wiita and Klu\'{z}niak-Lee pseudo-Newtonian gravitational
potentials $\Psi_{\mathrm g}$ are:
\begin{equation}
\Psi_{\mathrm{PW}}(r)=-\frac{GM_{\mathrm{BH}}}{r-r_{\mathrm{S}}}, \ \Psi_{\mathrm{KL}}(r)=-\frac{GM_{\mathrm{BH}}}{3 r_{\mathrm{S}}} (e^{3r_{\mathrm{S}}/r}-1).
\label{eq2}
\end{equation}
The initial disk is set up with a quasi-stationary result from the purely hydro-dynamical
(HD) simulations, which were obtained modifying the setup described in \cite{cem19}, with
the use of pseudo-Newtonian gravitational potentials. The \cite{KK00} analytical
solution was used as the initial disk set up for our HD simulations. In the magnetic
case, to the relaxed HD disk as an initial condition we add the magnetic field with
the vector potential described in \cite{zhus18} and \cite{bhup19}:
\begin{equation}
A_{\phi} = \left\{ \begin{array}{ll}
\frac{1}{2} r \sin \theta B_0 (\frac{r_{\mathrm {min}}}{R_0})^{m} & \mathrm{if\ r\leq r_{min}}\\
\frac{B_0}{R_0^m} \frac{(r \sin \theta)^{m+1}}{m+2} + 
\frac{B_0 r_{\mathrm {min}}^{m+2}}{R_0^m r \sin \theta}
(\frac{1}{2}-\frac{1}{m+2}) & \mathrm{if\ r>r_{min}}
\end{array} \right.
\label{eq3}
\end{equation}

\section{Results}

Snapshots of our preliminary results with Paczy\'{n}ski-Wiita and Klu\'{z}niak-Lee
potentials are shown in Fig. 1. Shown are the initial phases of the simulation,
before the disk relaxation to the action of magnetic field.
\begin{figure}
\centering
\includegraphics[width=.3\columnwidth,height=0.45\columnwidth]{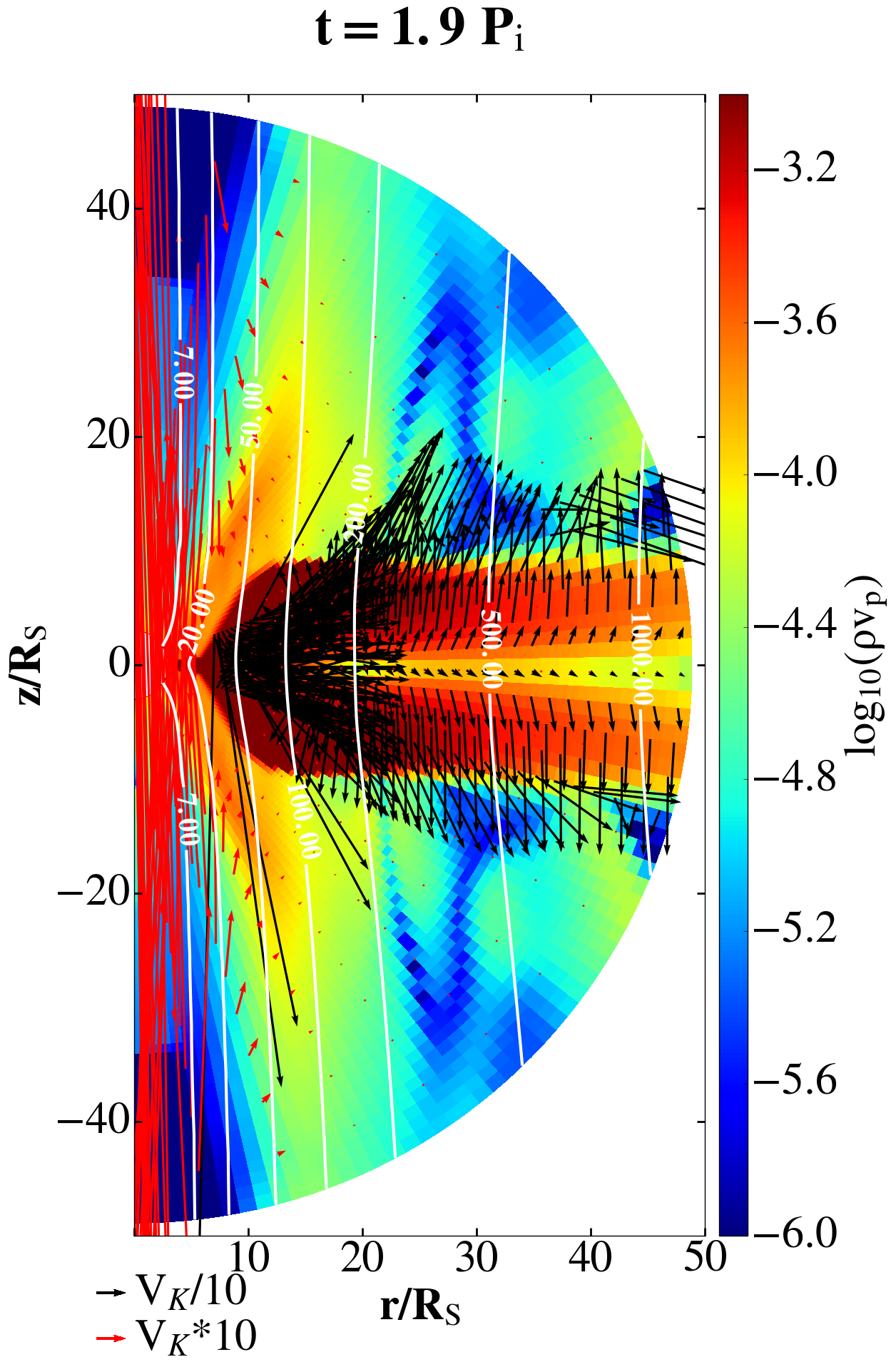}
\includegraphics[width=.3\columnwidth,height=0.45\columnwidth]{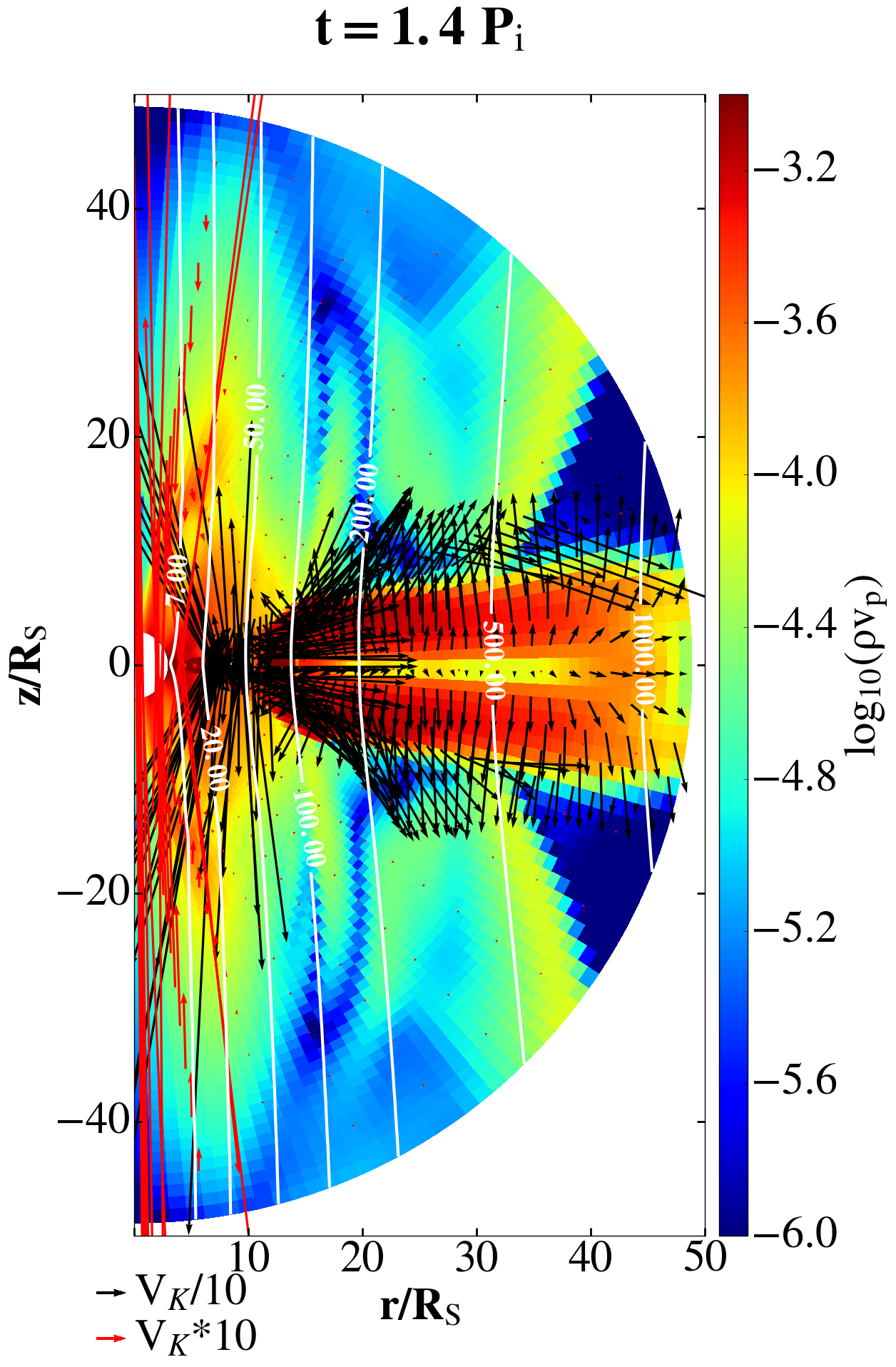}
\caption{Snapshot in a simulation with Paczy\'{n}ski-Wiita potential (left
panels) and Kluźniak-Lee potential (right panels). Poloidal momentum
flux ${\mathbf \rho v_{\mathrm p}}$ is shown in a
logarithmic color grading and a sample of magnetic field lines is shown with
solid white lines.  The poloidal velocity arrows are shown in
Keplerian velocity units at R$_{\mathrm i}$, with different
unit vector lengths in the disk and corona.}
\end{figure}
To better understand the disk dynamics, we computed the forces in each of
those cases (Fig. 2), projected along the magnetic field lines. The Lorentz
force is the strongest force acting on the disk in both cases, pulling it
apart from the disk equatorial plane. Such results will guide our further
work, search for the combination of physical parameters which would result
in the configuration with launching of the axial jet.
\begin{figure}
\includegraphics[width=.5\columnwidth,height=0.27\columnwidth]{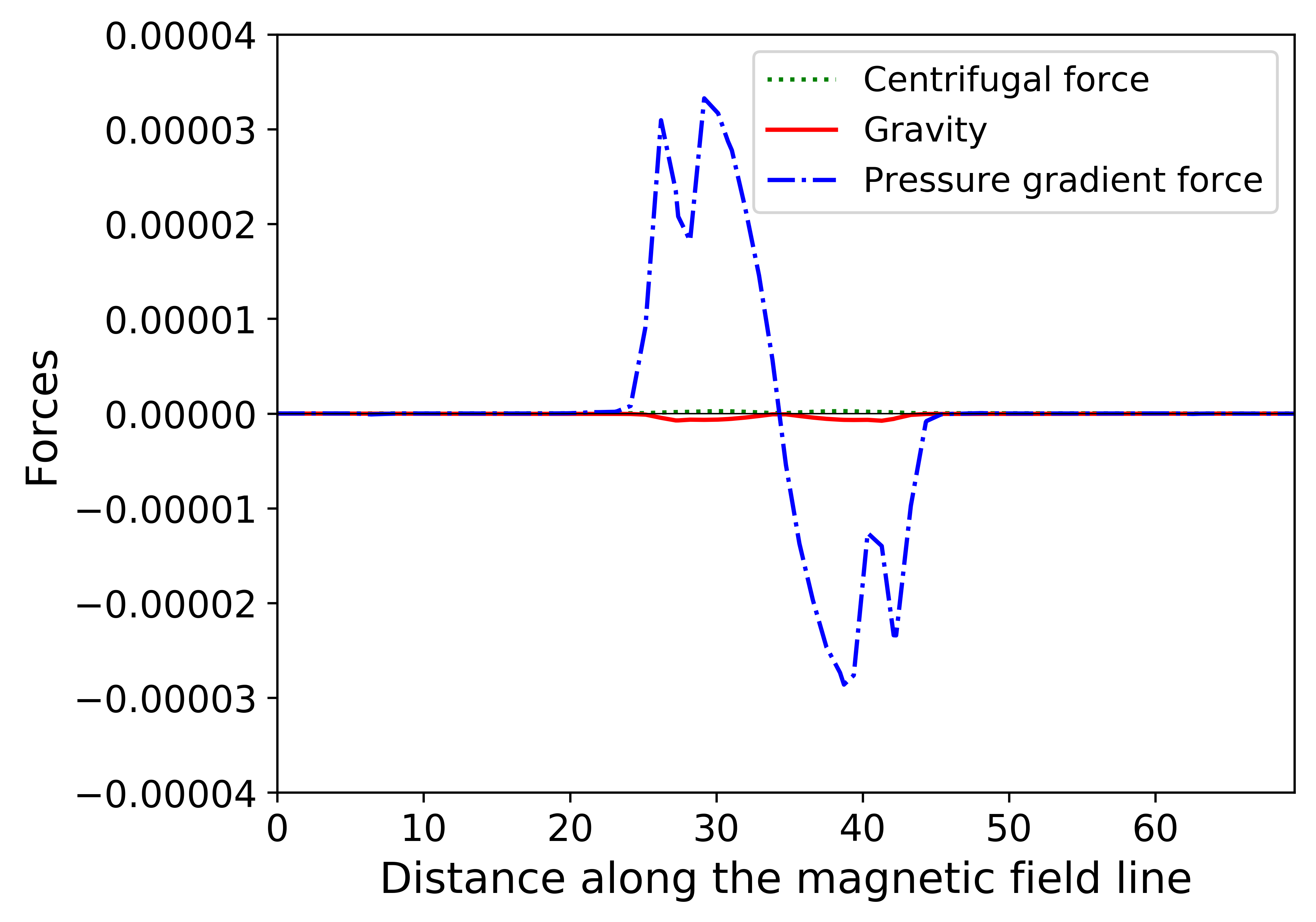}
\includegraphics[width=.5\columnwidth,height=0.27\columnwidth]{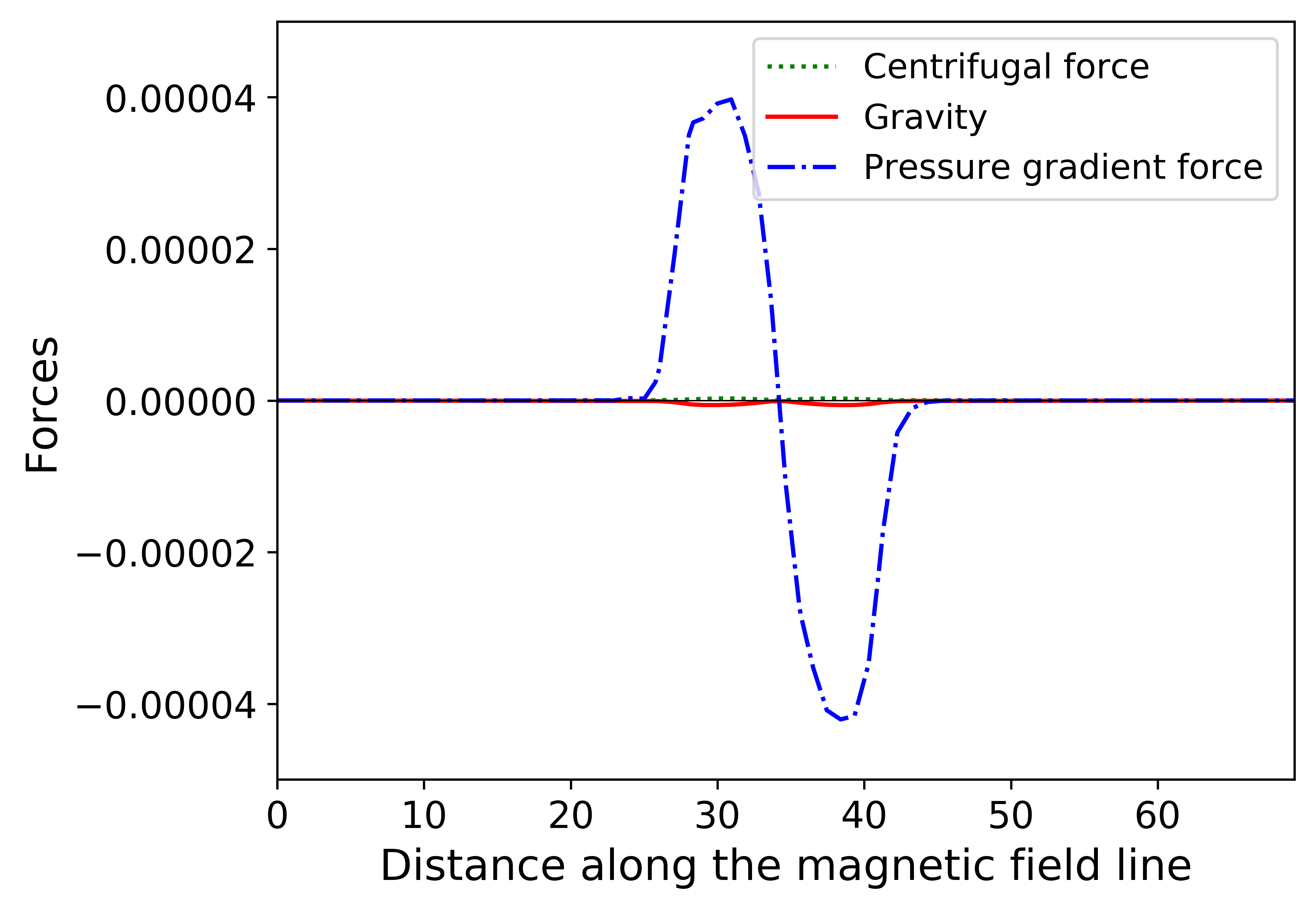}
\includegraphics[width=.5\columnwidth,height=0.27\columnwidth]{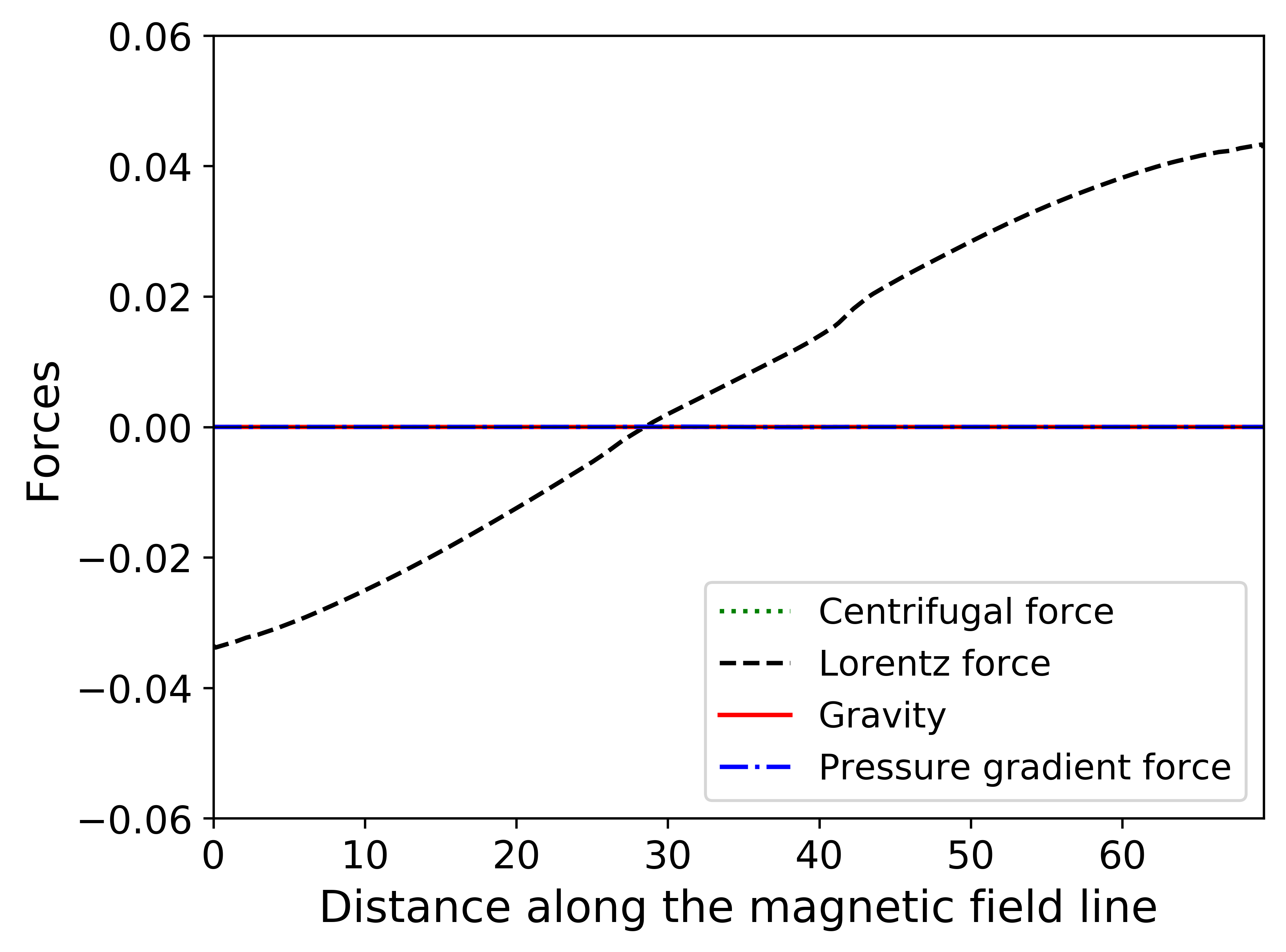}
\caption{Forces projected along a field line $\psi$=500, calculated from
the bottom to the top of the disk, in the cases with Paczyński-Wiita
potential (top left panel) and Kluźniak-Lee potential (top right panel).
Lorentz force is the same in both cases, and much larger than other forces,
so that we show it separately, with other forces ovelap at small values (bottom panel).}
\end{figure}

\section{Conclusions}
In our simulations with pseudo-Newtonian gravitational potentials, we are
investigating configurations of the magnetic field from which jets could be
launched from the vicinity of a supermassive black hole.  Starting from the
relaxed HD numerical solution, we study forces on the material in the disk
and its corona, searching for the configuration facilitating the launching
of outflows.

\acknowledgements{M. Jab\l o\'{n}ski's and others' work in Warsaw is funded
by the Polish NCN grant No.2013/08/A/ST9/00795. MČ developed the PLUTO
setup under ANR Toupies funding in CEA Saclay, France, and he also
acknowledges Croatian HRZZ grant IP-2014-09-8656.  We thank A. Mignone and
his team of contributors for the possibility to use the PLUTO code, and
ASIAA \& CAMK for use of their Linux clusters (PX, XL \& CHUCK).}

\bibliographystyle{ptapap}
\bibliography{jablonski}

\end{document}